 \documentstyle[floats,prl,aps,galley,psfig]{revtex}		

 \newlength{\figwidth}
 \newlength{\refspace}

\begin{document}
\draft


\Preprint				

\title{A Model for the Elasticity of Compressed Emulsions}
\author{Martin-D. Lacasse$^1$,
Gary S. Grest$^1$, Dov Levine$^2$,
T.G. Mason$^1$\cite{tom} and D.A. Weitz$^1$\cite{dave}}
\address{
$^1$Corporate Research Science Laboratories\\
Exxon Research and Engineering Co., Annandale, NJ 08801\\
$^2$Department of Physics, Technion, Haifa, 32000 Israel}
\date{December 13, 1995}

\maketitle%
{
\begin{abstract}
We present a new model to describe the unusual elastic properties of
compressed emulsions. The response of a single droplet under compression
is investigated numerically for different Wigner-Seitz cells. The response
is softer than harmonic, and depends on the coordination number of the
droplet. Using these results, we propose a new effective inter-droplet
potential which is used to determine the elastic response of a monodisperse
collection of disordered droplets as a function of volume fraction. Our
results are in excellent agreement with recent experiments. This suggests
that anharmonicity, together with disorder, are responsible for the
quasi-linear increase of $G$ and $\Pi$ observed at $\varphi_c$.
\end{abstract}
}
{
}

\narrowtext

Emulsions are materials with highly unusual elastic properties. They
consist of droplets of one fluid dispersed in a second fluid, with
interfaces stabilized by a surfactant. Despite being comprised solely of
fluids, they can be elastic solids, when droplets are compressed to a large
volume fraction, $\varphi$, by an osmotic pressure, $\Pi$. The origin of
this elasticity is the interfacial energy of the droplets. At low volume
fractions, their surface tension, $\sigma$, ensures that the droplets are
spherical in shape; however, at higher $\varphi$, the packing constraints
force the droplet shapes to deform, storing energy.
The application of a shear strain to a compressed emulsion causes
the droplets to deform further, increasing their surface
area, thereby storing elastic
energy~\cite{princenb,kraynik,bolton,morse,catesa}.

 \begin{figure}[hbt]
 \centerline{\psfig{figure=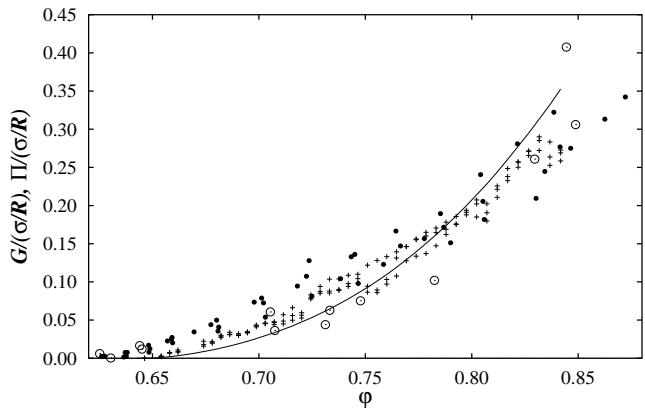,width=\figwidth}}
 \caption{Experimental values of
 the osmotic pressure ($\odot$) and
 elastic shear modulus ($\bullet$)
 of different emulsions with monodisperse droplets. The data are
 normalized by $\sigma/R$, allowing data
 from emulsions with different droplet sizes to be compared.
 Results predicted by the present model for the shear modulus ($+$)
 and the osmotic pressure (line) are also shown.
 \label{fig1}}
 \end{figure}

The experimentally measured~\cite{princena,masona} $\varphi$ dependence
of the static shear modulus $G$ is remarkable in several ways.
Typical data~\cite{masona} obtained for
monodisperse~\cite{bibetteb} silicone-oil-in-water emulsions are summarized
in Fig.~\ref{fig1}. The solid symbols show $G$ measured as a function
of $\varphi$ for several monodisperse emulsions of different droplet sizes.
The data are normalized by $\sigma/R$ where $R$ is the
radius of an undeformed droplet. The observed scaling by the Laplace
pressure $(2\sigma/R)$ confirms the
essential role of the interfacial energy. The magnitude of the scaled
modulus increases approximately linearly, varying as $G \sim
\varphi(\varphi-\varphi_c)$, where $\varphi_c$ is the volume fraction where
the droplets are first deformed. For monodisperse droplets, $\varphi_c
\approx 0.64$, the maximum volume fraction that monodisperse spheres can be
randomly packed~\cite{masona,berryman}; for polydisperse droplets
$\varphi_c$ is
larger~\cite{princena}, reflecting more efficient packing. Surprisingly,
the osmotic pressure required to compress the emulsion is very similar to
the shear modulus~\cite{masona,princenc}, as shown by the open
circles in Fig.~\ref{fig1}. This implies that the corresponding
longitudinal bulk elastic modulus, or the bulk osmotic modulus, $K =
\varphi d\Pi/d\varphi$, must differ significantly from the shear modulus
as it must have a much sharper onset at $\varphi_c$~\cite{masona}.

The behavior of the shear modulus of emulsions has been the subject
of various theoretical studies.
A first approach consists in considering a single droplet,
and to obtain the properties of a periodic structure,
assuming the strain remains affine~\cite{princenb,morse,catesa}.
While this can be done analytically in two dimensions, in three
dimensions it can only be done through approximations.
Results obtained from this approach consistently predict a
sharp rise of $G$ at $\varphi_c$, followed by a much slower
increase at larger $\varphi$.
This suggests that the effects of disorder may be responsible
for the quasi-linear rise of $G$ at $\varphi_c$ found experimentally.
The response to shear of disordered arrangements of droplets
has also been investigated. However, due to the additional complexity
this imposes, these calculations
have been restricted to two
dimensions~\cite{kraynik,bolton,hutzler,durian},
where disorder needs to be introduced through polydispersity.
Nevertheless, these
two-dimensional models show a different behavior than that
observed experimentally for disordered, three-dimensional
emulsions. In particular, two-dimensional
systems do not exhibit a smooth nearly linear increase
of $G$ near $\varphi_c$, although a smooth quadratic increase of $\Pi$
near $\varphi_c$ has recently been observed~\cite{hutzler}.

In this Letter, we present the first three-dimensional computer simulation
modeling the elasticity of a disordered emulsion. In addition,
we propose a new, more realistic inter-droplet potential.
It is based on numerical results obtained by
calculating the change in surface energy of
a single droplet as it is compressed using Brakke's
Surface Evolver (SE) software~\cite{brakke}.
Our model captures the essential physics of the droplet interactions
and their disordered, glassy packing. 
The droplets are soft, monodisperse spheres interacting
through a purely repulsive potential.
The droplet positions are allowed to relax upon application of shear.
We calculate the osmotic pressure and the shear modulus
as a function of volume fraction. We find that both the positional
relaxation of the droplets and the form of the potential are essential in
reproducing the experimental behavior of the shear modulus.

We first investigate in more detail the interaction between droplets.
When two droplets are forced together, their
spherical shapes are deformed and their
surfaces develop flattened facets at contact.
As a simplification, consider a droplet of radius $R$ compressed
between two parallel planes, each one located
at a distance $h$ from the center of the droplet.
Naively, for small
deformations $\delta\xi$, where $\xi = (R - h)/R$ is
a dimensionless measure of compression, the
resultant force, $F$, on the flat facets can be estimated by assuming that
the radius of the droplet, and hence the Laplace pressure, remains
unchanged.
We thus have $F \approx (2\sigma/R)\delta S$, where $\delta S$ is the area
of the flattened facet. To linear order in the deformation, $\delta S
\approx 2\pi R^2 \delta\xi$, so that $F \approx 4\pi\sigma R\delta\xi$.
Thus, the interaction between the facets of two droplets is often
taken as a strictly repulsive harmonic spring of spring constant $4\pi\sigma$
acting between the centers of the spheres~\cite{bolton,durian}.

The harmonic-spring potential, while appealing, ignores the details of the
response of the shape to deformation~\cite{morse,lacasse2},
and the possibility of
coupling between the different facets on each droplet. Thus, to determine
an improved potential, we investigated numerically the shape of a single
droplet confined within space-filling polyhedral cells. Using SE
we calculated the excess surface as the confinement is
increased, under the constraint of a fixed droplet volume.
As confining cells we investigated a rhombic dodecahedron
(f.c.c.), a truncated octahedron (b.c.c.) and a simple cube (s.c.). In all
cases, the compressions leave the center of mass unchanged, so that a
distance $h$ from the center can be defined.

 \begin{figure}[hbt]
 \centerline{\psfig{figure=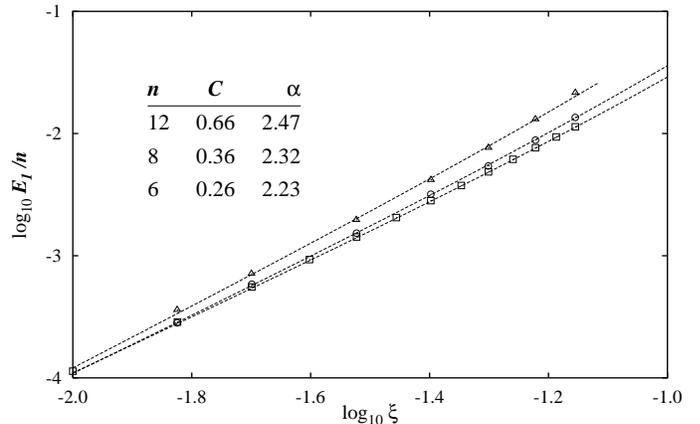,width=\figwidth}}
 \caption{Excess surface energy excess per facet under different
 Wigner-Seitz constraining cells. The lines are obtained from a fit
 to $E_1/n = C(\varphi/\varphi_c - 1)^\alpha$, where
 $\varphi/\varphi_c = (R/h)^3 = 1/(1-\xi)^3$.
 Curves are, from top to bottom:
 f.c.c.\ ($n=12,\ \varphi_c = \pi \protect\sqrt{2} /6$),
 b.c.c.\ ($n=8,\ \varphi_c = \pi \protect\sqrt{3} /8$),
 and s.c.\ ($n=6,\ \varphi_c = \pi/6$).
 Values of $C$ and $\alpha$ are accurate to 0.05.
 \label{fig2}}
 \end{figure}

For simplicity, we shall define $\sigma$ = 1, and $R$ = 1.
For a droplet of total surface area $A$, we can thus define the
excess energy as $E_1 = (A - 4\pi)$.
Figure~\ref{fig2} shows the calculated dependence of the
excess energy per
facet, $E_1/n$, on $\xi$ for different types of cells,
where $n$ represents the number of facets.
The data are well described by $[(R/h)^3 - 1]^\alpha$ as
shown by the dashed lines.  For small $\xi$,
this form reduces to a simple power law, $(3\xi)^\alpha$,
as is evidenced in Fig.\ref{fig2} by the asymptotically linear
behavior.
This functional form is also equivalent to
$(\varphi - \varphi_c)^\alpha$ for space-filling structures.
Fitting to $E_1/n = C[(R/h)^3 -1]^\alpha$, we find that
both the coefficient, $C$, and the
exponent, $\alpha$, depend on $n$,
as is evident from the data in Fig~\ref{fig2}.
Their values are shown in the table inserted in Fig.~\ref{fig2}.
The data selected for the fits
lie in the $\xi$ interval 0--7\%,
the upper value corresponding to $\varphi = 0.92$ for an f.c.c.\ lattice.
For the b.c.c.\ lattice we use $n = 8$
since second neighbors do not contribute to the energy for
$\varphi \lesssim 0.90$~\cite{lacasse2}.
Our results show explicitly that the response
of a droplet to compression is a non-local phenomenon: {\it the response
depends on the number of planes used to compress the droplet}.
One might expect that it also depends on the relative orientation
of these planes. However, comparison between results obtained
for iso-$n$ configurations, for instance, the pentagon
dodecahedron and the rhombic dodecahedron, suggests that the
distribution of the compressing planes on the droplet surface
has only a minor effect~\cite{lacasse2}. Thus, we expect
a similar functional form for $E_1$, even when the planes
are more randomly distributed as they would in a disordered
emulsion.

This behavior is different from the one encountered in two-dimensional
systems. A minimum free surface is characterized by a uniform pressure
or, equivalently by a uniform mean curvature. In two dimensions, the
surface is parameterized by only one radius of curvature and the minimum
free surface is always an arc of a circle. In this case, a harmonic
potential is a good approximation. By contrast, in three dimensions,
the response is always softer than harmonic since
$\alpha(n) > 2$ for all values of $n$ investigated here;
thus, the effective spring constant goes to zero as the distortion vanishes.
We note that, as proposed, a $n$-dependent power-law
leads to unphysical behavior for very small compressions, as the
energies for different $n$ should cross. However, this effect is too small
in magnitude to affect our results. The functional form we use
is a convenient way to mimic the response of a droplet over
the range of compression we investigate.

To describe the elastic properties of a disordered droplet packing,
we use a model which replaces the droplets by soft spheres
which interact with their nearest neighbors through central-force
potentials that reflect the behavior of the facets.
However, since we found that the energy curve of each
contact is a function of the droplet coordination number,
the interaction energy should be obtained by balancing
the forces at each contact, using the values of $n$ for each droplet.
To make the computation tractable, we determine the average coordination
number, $\bar{n}$, of the whole configuration of droplets, and use a single
effective potential for that configuration.
This is a reasonable approximation
given the rather narrow distribution of $n$
for our monodisperse systems.
Thus, we use the SE results to define a repulsive,
central-force inter-droplet potential,
\begin{equation}
\label{potential}
U(d) = \left\{
\begin{array}{ll}
2C \left[(\frac{2R}{d})^3 - 1\right]^\alpha, & (d < 2R)\\
0, & (d \geq 2R)
\end{array}
\right.
\end{equation}
where $d = 2h$ is the distance between the droplet centers,
and the factor 2 accounts for the two facets on the interacting pair.
During the simulation, the exponent $\alpha(\bar{n})$ and the prefactor
$C(\bar{n})$ are estimated from cubic interpolations of the values
shown in Fig.\ref{fig2}.
The value of $\bar{n}$ is observed to
increase from $\bar{n}_c \approx 6$ at $\varphi_c \approx 0.64$ to
$\bar{n} \approx 10$ at $\varphi = 0.84$.
We finally note that central-force potentials
can only include compressional distortion;
nevertheless we use them to describe shear
distortions as well.

The relaxation algorithm entails
the minimization of the 3N-dimensional function,
\begin{equation}
E_N = \sum_{i,j>i}^N U(d_{ij}),
\end{equation}
where $d_{ij}$ is the distance between point-particles $i$ and $j$. We use
a conjugate-gradient (CG) method coupled to Brent line-minimization
method~\cite{brent}. This process represents a damped relaxation of a
system of interacting droplets, and therefore the masses of the particles
are not necessary.
In order to find only a local minimum,
our version of CG is written so that
each line-minimization search interval is defined
using the result of the line-minimization
along the previous conjugate direction.
The evaluation of the potential
benefits from techniques borrowed from molecular dynamics, such as a Verlet
table~\cite{allen}. The algorithm is designed
to return after a minimum number of iterations {\it and}\/
when new values for the energy differ by less than
$\delta E_N/E_N \leq 10^{-7}$.
For each relaxed configuration, the energy, the average coordination
number, and the pressure are measured. The coordination number is derived
from the sum of all pairs contributing at least a small $\epsilon$ to
the energy and values for the pressure in each Cartesian direction
(diagonal elements of the stress tensor)
are obtained from the virial~\cite{allen}.

We first construct a random distribution of $N$ monodisperse
particles of radius $R$ in a cubic
container with periodic boundary conditions.
Typical runs start from a configuration
prepared at the desired volume fraction.
Systems that were slowly compressed from
$\varphi < \varphi_c$ to 
$\varphi > \varphi_c$, and random
configurations built and relaxed at $\varphi > \varphi_c$ were
found to have similar elastic properties.
Thus, we usually start at
$\varphi \approx 0.67 > \varphi_c$ in order to avoid
the long computational time required to relax a configuration
at $\varphi_c$. The system is then compressed and relaxed
in small step increments up to
$\varphi \approx 0.84$. From this value,
the cubic container is sheared using
isochoric uniaxial strains
wherein, say, the $z$ axis is stretched by
a factor $\lambda \gtrsim 1$
and the perpendicular plane is compressed by
$\lambda^{-\frac{1}{2}}$.
The shear modulus is obtained from the
excess energy density as a function of the extension ratio
$\lambda$~\cite{treloar},
\begin{equation}
E_N/V = E_N^o/V + \frac{1}{2} G
(\lambda^2 + 2/\lambda - 3)
\label{shearit}
\end{equation}
where $E_N^o$ is the excess energy of the unstrained system.
We perform several strain cycles
to allow for relaxation and to verify reproducibility. This shearing
process is performed along the three Cartesian directions.
The system is then expanded in a small step increment,
relaxed, and the shearing procedure is repeated.
System sizes of at least $N = 10^3$ were used to avoid
undesired relaxation into an f.c.c.\ structure at large strains.

The calculated $\varphi$ dependence of the shear modulus is shown by the
plus symbols in Fig.~\ref{fig1}, while the osmotic pressure is shown by
the solid line, for a system of $N = 10^3$. Interestingly, the data
for $\Pi$ shows self-averaging and contains much less fluctuations
that that for $G$. This suggests
that there is an effective larger correlation length that determines
the modulus and limits the averaging possible in our finite systems.
Nevertheless, remarkably good agreement with
the experimental data is obtained. Both the magnitudes and the $\varphi$
dependencies are correctly reproduced for both $G$ and $\Pi$.

 \begin{figure}[hbt]
 \centerline{\psfig{figure=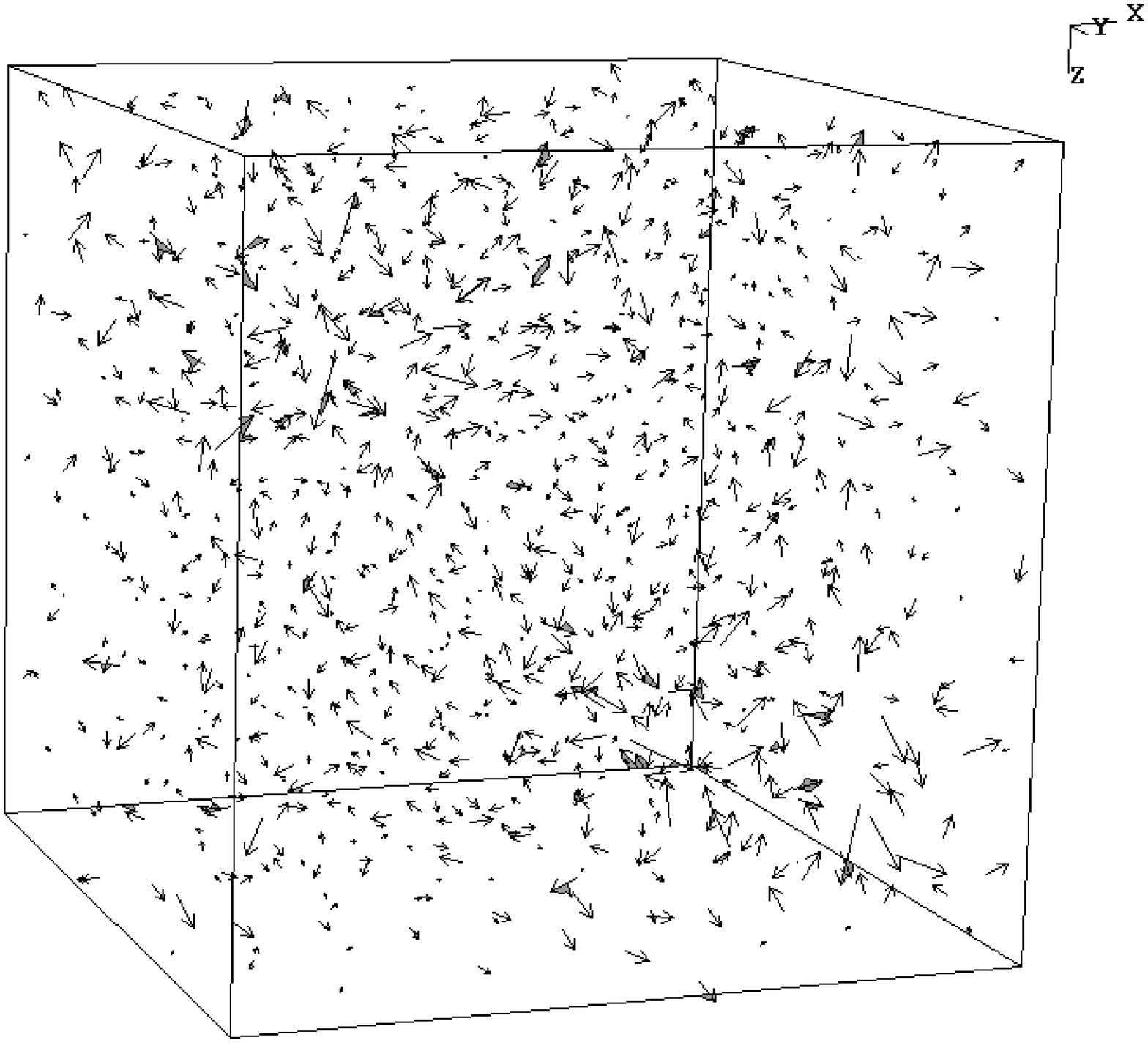,width=\figwidth}}
 \caption{An isochoric uniaxial shear strain
 of $\lambda = 1.0006$ is applied to a $N = 10^3$ system. Each
 arrow, magnified 250 times,
 represents the droplet relaxation from an affine, non-relaxed
 strain. $\varphi = 0.80$.
 \label{fig3}}
 \end{figure}

The simulation also provides physical insight about the origin of
the behavior of the shear modulus of emulsions. There are two essential
effects. The first is the positional relaxation of the droplets.
We illustrate this graphically in Fig.~\ref{fig3}:
we subtract the actual motion of the droplets from the affine motion caused
by a strain associated with $\lambda = 1.0006$,
and use arrows at the center of each droplet to
signify the direction and magnitude of this difference. The length of
each arrow has been increased by 250 to make the motion with such a small
strain observable. The droplet motion is clearly not affine; moreover, the
difference appears random in direction. The importance of this
non-affine motion is reinforced by calculating the shear modulus for a
strictly affine (unrelaxed) motion; it increases by a factor of about 3 for all
$\varphi$. We emphasize, however, that the non-affine motion of the droplets
does not result in large scale rearrangements of their positions. We found
that only a very small fraction of the droplets change neighbors
during shear. Thus, the non-affine motion results from localized
relaxation of the droplet positions~\cite{liu}.
The second essential feature is the anharmonicity of the potential.
We performed similar simulations using a harmonic potential and
obtained qualitatively different
results~\cite{lacasse} for $G$, which
exhibited a significantly steeper rise near
$\varphi_c$, similar to what is observed in harmonic
two-dimensional systems~\cite{bolton,durian}.
By contrast, the behavior of $\Pi$,
is not as sensitive to the potential.

Finally, we note that the average coordination number
was found to increase continuously above $\varphi_c$,
consistent with a power law increase
$(\bar{n} - \bar{n}_c) \sim (\varphi - \varphi_c)^\frac{1}{2}$
surprisingly the same functional form as the one observed in
computer simulations in two dimensions~\cite{durian}.
In the present picture, the increase of $\bar{n}$ not only
creates more contacts capable of storing energy, but
also causes the energy of the existing contacts
to increase because of the increase of the number of facets
around each droplet. However, this has only a minor effect as confirmed
by simulations using fixed $C$ and $\alpha$.

These results suggest an explanation for the origin of the
surprising elasticity of emulsions; moreover,
they will likely have broader significance. They are immediately
applicable to foams, since the compressibility of the gas in the bubbles
is typically much less than the Laplace pressure. More generally,
emulsions are an example of a material with strictly repulsive
interactions which is nevertheless a solid when confined by an osmotic
pressure. The elasticity of these packings is significantly different
than that of more traditional materials, and the results presented here
should form the basis for developing a more comprehensive description
of these fascinating materials.

We thank Shlomo Alexander for useful discussions, {\it Le Fonds
FCAR du Qu\'ebec}\/ and
the U.S.-Israel Binational Science Foundation
for financial support.



 \end{document}